\documentclass{appolb}%
\usepackage{amssymb}
\usepackage[usenames,dvipsnames]{color}
\usepackage[colorlinks=true,linkcolor=darkblue,citecolor=darkblue,urlcolor=darkblue]%
{hyperref}
\usepackage{graphicx}
\usepackage{makecell}
\usepackage{bm}
\usepackage{amsmath}%
\usepackage{amsfonts}
\pdfoutput=1
\definecolor{darkblue}{RGB}{0,0,196}
\begin{document}

\title{Irrelevance of $f_{0}(500)$ in bulk thermal properties 
\thanks{Talk presented at XI Workshop on Particle Correlations and Femtoscopy, 3-7 November 2015, Warsaw, Poland.}
}
\author{Francesco Giacosa$^{1,2}$, Viktor Begun$^{1}$, \\and Wojciech Broniowski$^{1,3}$
\address{$^{1}$Institute of Physics, Jan Kochanowski University, PL-25406,
Kielce, Poland
\\$^{2}$Institute for Theoretical Physics,
Goethe-University, D-60438, Frankfurt am Main, Germany
\\$^{3}$The
H. Niewodnicza\'{n}ski Institute of Nuclear Physics, Polish Academy of
Sciences, PL-31342, Krak\'{o}w, Poland}
}
\maketitle

\begin{abstract}
We discuss why the scalar-isoscalar resonance $f_{0}(500)$ should in practice
not be included in thermal models describing the freeze-out of heavy-ion
collisions. Its contribution into pion multiplicities is in principle relevant
because it is light and it decays only into pions. However, it is cancelled
to a very good numerical precision by the non-resonant scalar-isotensor
repulsion among pions. Our approach is an application of a well-known theorem
relating spectral function to phase shifts. The numerical results are solely
based on pion-pion scattering data and thus model independent.

\end{abstract}


\section{Introduction}

The scalar-isoscalar resonance $f_{0}(500)$ is now firmly established
\cite{pelaezrev}. The Particle Data Group (PDG) reports the position of its
pole in the range $\left(  400-550\right)  -i(200$-$350$)~\cite{pdg}.
Investigations based on dispersive analysis show even smaller errors: $\left(
400\pm6_{-13}^{+31}\right)  -i(278\pm6_{-43}^{+34}$) in Ref.~\cite{caprini}
and $(457_{-13}^{+14})-i(279_{-7}^{+11})$ in Ref.~\cite{kaminski}.

The resonance $f_{0}(500)$ is the lightest scalar state; moreover, it decays
only into pions. Then, one is lead to think that $f_{0}(500)$ is important for
the determination of pion multiplicities in thermal models for relativistic
heavy-ion collisions (see e.g. Refs.~\cite{wfbook,giorgio} and refs. therein).
Indeed, a simple inclusion of $f_{0}(500)$ as a Breit-Wigner resonance would
lead to a sizable increase (about 3-5\% \cite{andronic}) of pions. However, in
these proceedings (based on the findings of Ref. \cite{our}), we show that
this conclusion is not correct. In fact, when the repulsion of pion-pion
interaction in the scalar-isotensor channel is properly taken into account
using the formalism developed in Refs. \cite{dashen,weinhold,denis}, the
effect of $f_{0}(500)$ cancels to a very good numerical accuracy. We show this
cancellation in a model independent way, since the only input is given by
well-known pion-pion scattering data in these two scalar channels.

\ As a net result, one can neglect in thermal models both the scalar-isoscalar
attraction due to $f_{0}(500)$ and the non-resonant scalar-isotensor repulsion.

\section{Cancellation of $f_{0}(500)$}

A successful description of hadron emissions at the freeze-out of relativistic
heavy-ion collisions is achieved with the help of thermal models.\ For
simplicity, we restrict our presentation to a gas which includes stable pions
($I=0,$ $J^{PC}=0^{-+},$ where $I$ stays for isospin, $J$ for the total spin,
and $P$ and $C$ for parity and charge-conjugation), the $\rho$-resonance
($I=1$, $J^{PC}=1^{--}$ ), the resonance $f_{0}(500)$ ($I=0,$ $J^{PC}=0^{++}%
$), and the non-resonant contribution of the repulsion in the $I=2,$
$J^{PC}=0^{++}$ channel. (Other contributions with different $I$ and $J^{PC}$
correspond to heavier mesons and are neglected here.)

The logarithm of the partition function $Z$ is given by the sum of
contributions of all channels:
\[
\ln Z=\ln Z_{\pi}+\ln Z_{(1,1^{--})}+\ln Z_{(0,0^{++})}+\ln Z_{(2,0^{++}%
)}\text{ .}%
\]
All other thermodynamic quantities follow: $P=-T\ln Z/V,$ $\varepsilon
=-\partial_{\beta}\ln Z/V,$ etc. For what concerns stable pions (we do not
include chemical potentials), one has
\[
\ln Z_{\pi}=3V\int_{p}\ln\left[  1-e^{-\frac{\sqrt{\vec{p}^{2}+M_{\pi}^{2}}%
}{T}}\right]  ^{-1},\text{ }\int_{p}\!=\int\!\frac{d^{3}p}{(2\pi)^{3}}\text{
,}%
\]
where $V$ is the volume, $\vec{p}$ the pion momentum, $M_{\pi}$ the pion mass,
and the factor $3$ the isospin degeneracy. Following Refs.
\cite{dashen,weinhold} we can express the contribution in the channel $I=1,$
$J^{PC}=1^{--}$ as:
\begin{equation}
\ln Z_{(1,1^{--})}=3\cdot3\int_{0}^{\Lambda_{1}}\!\!\!\!\!\!\text{ }%
dm\frac{d\delta_{(1,1)}(m)}{\pi dm}\,\int_{p}\ln\left[  1-e^{-\frac
{\sqrt{p^{2}+m^{2}}}{T}}\right]  ^{-1}\text{ ,} \label{rho}%
\end{equation}
where $\delta_{(1,1)}$ is the measured $\pi\pi$ phase-shift as function of
$m=\sqrt{s}.$ We set $\Lambda_{1}=1$ GeV as maximal energy in the integral,
then only the $\rho$-meson is present in this range. The spectral function of
the $\rho$-meson is related to the phase-shift by \cite{dashen,weinhold}:
\begin{equation}
d_{\rho}(m)=\frac{d\delta_{(1,1)}(m)}{\pi dm}\text{ .}%
\end{equation}
Thus, one can take into account the $\rho$-meson in a thermal gas in a model
independent way by introducing the well-known scattering data in Eq.
(\ref{rho}). For a small width $d_{\rho}(m)$ can be well approximated by a
Breit-Wigner function, $d_{\rho}(m)\simeq\frac{\Gamma}{2\pi}\left[
(m-M_{\rho})^{2}+\Gamma^{2}/4\right]  ^{-1}$ , and, in the limit of zero
width, one correctly obtains $d_{\rho}(m)=\delta(m-M_{\rho}).$ Thus, the
example of the $\rho$-meson shows quite general features of a thermal gas. The
approximation of using a Breit-Wigner expression -typically used in practice- emerges.

We now turn to the main topic of the present work. For $I=J=0,$ the
contribution of $f_{0}(500)$ is included in the integral%
\begin{equation}
\ln Z_{(0,0^{++})}=\int_{0}^{\Lambda_{0}}\!\!\!\!\!\!dm\frac{d\delta_{(0,0)}%
}{\pi dm}\int_{p}\ln\left[  1-e^{-\frac{\sqrt{p^{2}+m^{2}}}{T}}\right]  ^{-1}%
\end{equation}
where $\Lambda_{0}\simeq0.8$ GeV (far above the average mass of $f_{0}(500)$
but below $f_{0}(980),$ which is not considered here). The spectral function
of the $f_{0}(500)$ is given by $d_{f_{0}(500)}(m)=\frac{1}{\pi}%
d\delta_{(0,0)}/dm$. The form of $d_{f_{0}(500)}(m)$ is far from being a
Breit-Wigner \cite{our} and is even not normalized to unity. This is in
agreement with the fact that the resonances $f_{0}(500)$ is not the chiral
partner of the pion and is not a quark-antiquark field \cite{pelaezrev} (the
chiral partner of $\pi$, the `true' $\sigma$ of linear Sigma Models, should be
identified with the heavier scalar resonance $f_{0}(1370)$ \cite{dick}).

As a last step, we consider the joint contribution of both $I=0$ and $I=2$
channels:%
\begin{equation}
\ln Z_{(0,0^{++})}+\ln Z_{(2,0^{++})}=\int_{0}^{\Lambda_{0}}\!\!\!\!\!\!\text{
}dm\left[  \frac{d\delta_{(0,0)}}{\pi dm}+5\frac{d\delta_{(2,0)}}{\pi
dm}\right]  \,\int_{p}\ln\left[  1-e^{-\frac{\sqrt{p^{2}+m^{2}}}{T}}\right]
^{-1}%
\end{equation}
where the factor $5$ in front of $d\delta_{(2,0)}/dm$ is the degeneracy
$2I+1.$ \textit{Data} on pion-pion scattering show the following peculiar fact
\cite{our}:
\begin{equation}
\frac{d\delta_{(0,0)}}{\pi dm}+5\frac{d\delta_{(2,0)}}{\pi dm}\simeq0\text{
for }2M_{\pi}\leq m\lesssim0.8\text{ GeV,}%
\end{equation}
which is valid to a very good numerical accuracy. Then $\ln Z_{(0,0^{++})}+\ln
Z_{(2,0^{++})}\simeq0$! The contribution of $f_{0}(500)$ cancels.

\section{Conclusions}

In these proceedings we have shown that the resonance $f_{0}(500)$ can in
practice be neglected in all isospin-averaged quantities of a thermal hadronic
gas, e.g. for pion multiplicities. Then, the proton-to-pion puzzle becomes
even stronger, leaving other explanations open~\cite{begun}. On the other
hand, in correlation studies of pion-pair production the cancellation does not
occur, hence $f_{0}(500)$ may play a relevant role~\cite{hiller}.

A similar cancellation occurs for the not yet confirmed $K_{0}^{\ast}(800)$
($I=1/2,J^{PC}=0^{++},$ e.g. Ref. \cite{milena} and refs. therein), whose
contribution is (only partly) compensated by the $I=3/2,J^{PC}=0^{++}$ channel
\cite{our,redlichk}.

\bigskip

\textbf{Acknowledgments:} We thank R. Kami\'{n}ski, W. Florkowski, J. Pelaez,
G. Pagliara, V. Mantovani Sarti, T. Wolkanowski for useful discussions. This
research was supported by the Polish National Science Center, Grant DEC-2012/06/A/ST2/00390.

\end{document}